\documentclass[8.5pt,twoside,twocolumn]{article}
\usepackage[super,sort&compress,comma]{natbib}
\usepackage[version=3]{mhchem}
\usepackage{balance}
\usepackage{times,mathptm}
\usepackage{sectsty}
\usepackage{graphicx}
\usepackage{lastpage}
\usepackage[format=plain,justification=raggedright,singlelinecheck=false,font=small,labelfont=bf,labelsep=space]{caption}
\usepackage{fancyhdr}
\begin{document}

\thispagestyle{plain} \fancypagestyle{plain}{
\renewcommand{\headrulewidth}{1pt}}
\renewcommand{\thefootnote}{\fnsymbol{footnote}}
\renewcommand\footnoterule{\vspace*{1pt}%
\hrule width 3.4in height 0.4pt \vspace*{5pt}}
\setcounter{secnumdepth}{5}

\makeatletter
\renewcommand\@biblabel[1]{#1}
\renewcommand\@makefntext[1]%
{\noindent\makebox[0pt][r]{\@thefnmark\,}#1}
\makeatother
\renewcommand{\figurename}{\small{Fig.}~}
\sectionfont{\large}
\subsectionfont{\normalsize}

\fancyfoot{}
\fancyfoot[RO]{\footnotesize{\sffamily{1--\pageref{LastPage} ~\textbar  \hspace{2pt}\thepage}}}
\fancyfoot[LE]{\footnotesize{\sffamily{\thepage~\textbar\hspace{3.45cm} 1--\pageref{LastPage}}}}
\fancyhead{}
\renewcommand{\headrulewidth}{1pt}
\renewcommand{\footrulewidth}{1pt}
\setlength{\arrayrulewidth}{1pt}
\setlength{\columnsep}{6.5mm}
\setlength\bibsep{1pt}

\twocolumn[
  \begin{@twocolumnfalse}
\noindent\LARGE{\textbf{Arrays of carbon nanoscrolls as
deep-subwavelength magnetic metamaterials$^\dag$}} \vspace{0.6cm}

\noindent\large{\textbf{Vassilios
Yannopapas,$^{\ast}$\textit{$^{a}$} Marilena
Tzavala,\textit{$^{a}$} and Leonidas
Tsetseris\textit{$^{a}$}}}\vspace{0.5cm}

\noindent\textit{\small{\textbf{Received Xth XXXXXXXXXX 20XX, Accepted Xth XXXXXXXXX 20XX\newline
First published on the web Xth XXXXXXXXXX 200X}}}

\noindent \textbf{\small{DOI: 10.1039/b000000x}}
 \end{@twocolumnfalse} \vspace{0.6cm}

  ]

\noindent\textbf{We demonstrate theoretically that an array of
carbon nanoscrolls acts as a hyperbolic magnetic metamaterial in
the THz regime with genuine subwavelength operation corresponding
to wavelength-to-structure ratio of about 200. Due to the low
sheet resistance of graphene, the electromagnetic losses in an
array of carbon nanoscrolls are almost negligible offering a very
sharp magnetic resonance of extreme positive and negative values
of the effective magnetic permeability. The latter property leads
to superior imaging properties for arrays of carbon nanoscrolls
which can operate as magnetic endoscopes in the THz where magnetic
materials are scarce. Our optical modelling is supplemented with
ab initio density-functional calculations of the self-winding of a
single layer of graphene onto a carbon nanotube so as to form a
carbon nanoscroll. The latter process is viewed as a means to
realize ordered arrays of carbon nanoscrolls in the laboratory
based on arrays of aligned carbon nanotubes which are nowadays
routinely fabricated.}
\section*{}
\vspace{-1cm}
\footnotetext{\dag~Electronic Supplementary Information (ESI) available:
[details of any supplementary information available should be included here]. See DOI: 10.1039/b000000x/}


\footnotetext{\textit{$^{a}$~Department of Physics, National
Technical University of Athens, GR-15780 Athens, Greece. Fax: +30
210 7723025; Tel: +30 210 7721481; E-mail: vyannop@mail.ntua.gr}}



Metamaterials are artificial materials which exhibit response
characteristics that are not observed in the individual responses
of its constituent materials such as artificial magnetism,
negative refractive index, near-field amplification, cloaking and
optical illusions. \cite{shalaev_book} The basic functionalities
of metamaterials stem from the occurrence of electric/ magnetic
resonances wherein the electromagnetic field (EM) field is
strongly localized within subwavelength volumes. Their magnetic
response is associated with the induction of strong currents in
illuminated metamaterials. These strong currents, in turn, can
lead to strong paramagnetic (permeability $\mu>1$) and diamagnetic
behaviour (permeability $\mu<1$ or even $\mu<0$) in the
near-infrared and optical regions where such a response is not met
in naturally occurring materials. Magnetic activity in these
regions of the EM spectrum is of great technological importance
since it allows for the realization of devices such as compact
cavities, adaptive selective lenses, tunable mirrors, isolators,
converters, optical polarizers, filters, and phase shifters.
\cite{solymar}

The basic requirement for defining a given artificial EM structure
as a metamaterial is its subwavelength nature, i.e., the operating
wavelength being much larger than the characteristic length,
(e.g., period) of the structure. The higher the
wavelength-to-structure ratio the most efficient the operation of
metamaterials is. Namely, undesirable effects related with the
corresponding effective-medium parameters, i.e., the effective
electric permittivity $\epsilon_{eff}$ and magnetic permeability
$\mu_{eff}$, are sufficiently mitigated. Such effects are, for
example, the wavevector dependence due to the spatial
inhomogeneity \cite{tserkezis_09} or the anti-resonance behaviour
and the concomitant unnatural negative imaginary parts for
$\epsilon_{eff}$ and/ or $\mu_{eff}$. \cite{koschny_05}

To the best of our knowledge, the deepest subwavelength
metamaterial design reported so far has been the so-called `Swiss
Roll' array, \cite{pendry_99} a two-dimensional (2D) array of
rolled-up meta-atoms consisting of several turns of a metallic
(e.g., copper) laminate wound onto a central former.
\cite{wiltshire_01,wiltshire_03,wiltshire_07} Swiss Roll arrays
operate in the MHz regime and are principally used as endoscopes
for the magnetic field in magnetic resonance imaging.
\cite{wiltshire_01,wiltshire_03} The wavelength-to-structure ratio
$\lambda / a$  ($a$ being the unit cell size and $\lambda$ the
free-space wavelength) for a Swiss Roll metamaterial operating at
21.5~MHz is  $\lambda / a > 1000$, orders of magnitude higher than
conventional metamaterial designs based on split-ring resonators
where typically $\lambda / a \approx 5$. However, Swiss-Roll
meta-atoms with a large number of turns are very hard to
miniaturize in the infrared (IR) and visible regimes due to their
elaborate shape. On the contrary, metamaterial designs based on
split-ring resonators have been scaled down to size as small as
200~nm. \cite{enkrich,soukoulis_11}

In this work we propose a miniaturized version of the 'Swiss Roll'
metamaterial design based on carbon nanoscrolls (CNS). We show
that 2D arrays of CNS behave as THz hyperbolic magnetic
metamaterials with a hyperbolic photon dispersion relation that
allows canalisation of the near-field (evanescent) components of a
light field. \cite{cortes,drachev} A recent study \cite{iorsh}
proposed the formation of an electric hyperbolic metamaterial
using a multi-layered stack of alternating graphene layers and
dielectric slabs. The CNS-based metamaterial design proposed here
is {\it magnetically} active in the THz regime where such
materials are particularly scarce, with extreme subwavelength
operation corresponding to wavelength-to-structure ratio $\lambda
/ a$ as high as 200.  This design is ideally suited for magnetic
endoscope in the THz regime, as well as for negative-index
metamaterial when combined with a negative $\epsilon_{eff}$
structure.

\begin{figure}[h]
\centering
\includegraphics[width=8cm]{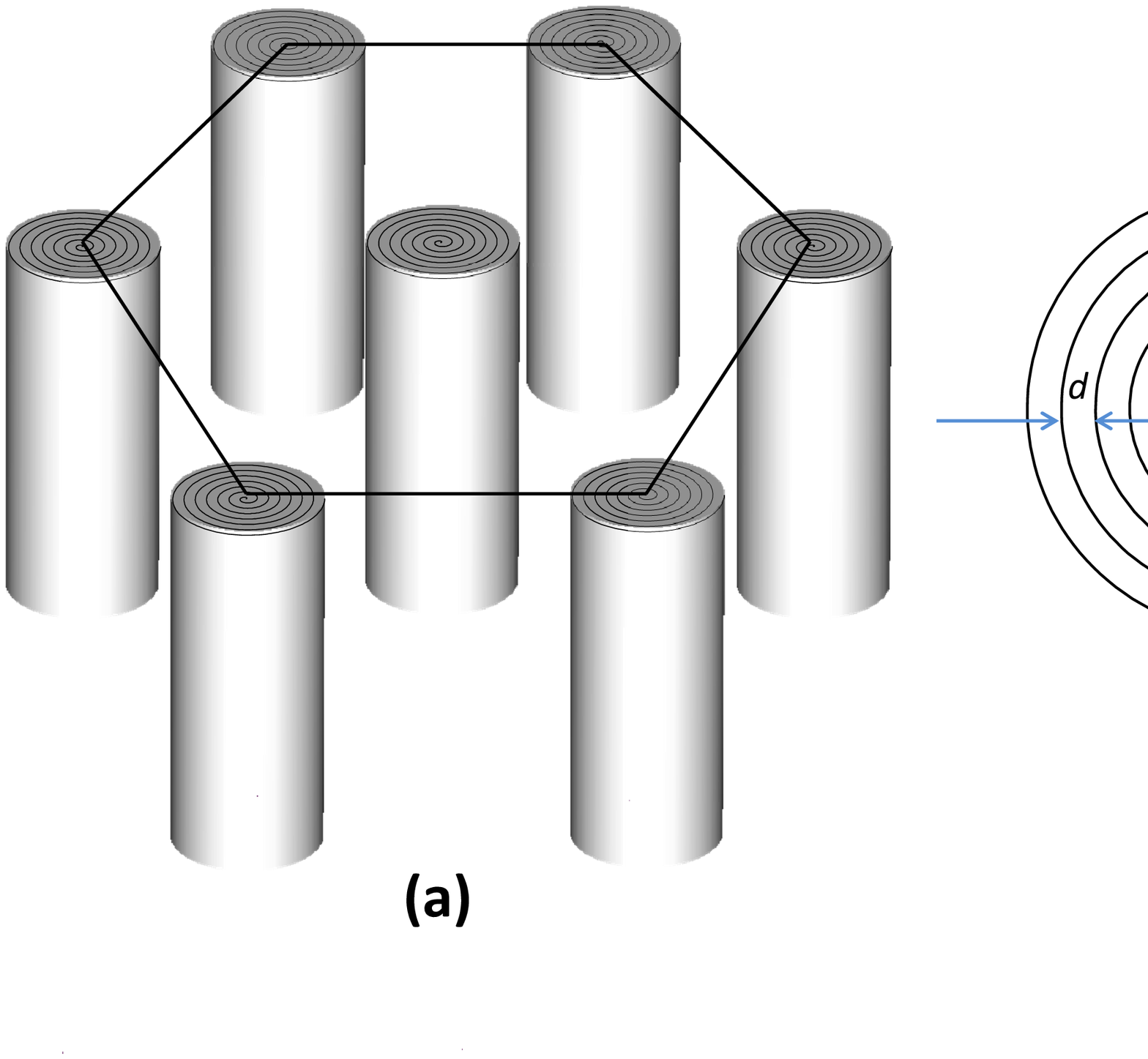}
\caption{(a) Hexagonal array of 'Swiss rolls' and (b) an
archimedean spiral} \label{fig1}
\end{figure}

A carbon nano-scroll is formed when a graphene sheet rolls into
the spiral geometry of Fig.~\ref{fig1}. Experimental
evidence~\cite{amelinckx_sci95} suggests that multi-walled carbon
nanotubes may appear both in the form of rolled graphene sheets
and in the more familiar geometry of concentric cylinders.
Theoretically, it has been shown~\cite{braga_nanol04} that
self-folding of a graphene layer into a CNS is an exothermic
transformation, but is hindered in the initial stages because of
an energy barrier. In the presence of a carbon nanotube, however,
the barrier may become small,\cite{zhang_apl10} enabling the
facile wrapping of a papyrus-like CNS around the tube. In the last
part of the following description of results, we will analyze the
atomic-scale details of CNS wrapping based on first-principles
calculations.

We assume an hexagonal lattice of CNS of lattice constant $a$ -
see Fig.~\ref{fig1}a. Figure~\ref{fig1}b shows the top view of a
CNS in the form of an archimedean spiral where the distance
between two successive turns is $d=0.35$~nm (the same as the
distance between two successive graphite layers) and the overall
radius of the CNS is $r=Nd$ where $N$ is the number of turns of
the CNS.

\begin{figure}[h]
\centering
\includegraphics[width=8cm]{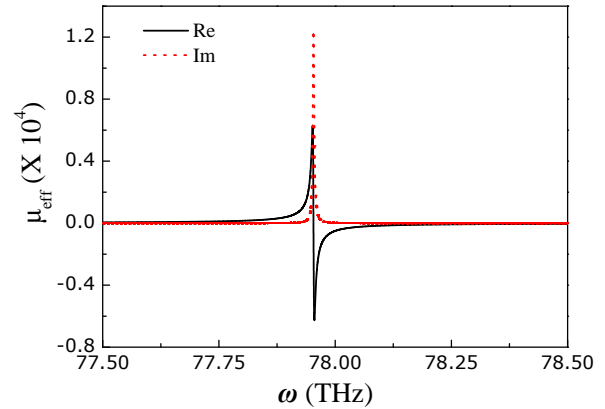}
\caption{Effective magnetic permeability of a 2D array of undoped
carbon nanoscrolls (CNS) with 60\% surface coverage. Each CNS
consists of 50 turns of graphene resulting in a resonance
frequency $\omega_0$=77.95~THz.} \label{fig2}
\end{figure}

\begin{figure}[h]
\centering
  \includegraphics[width=8cm]{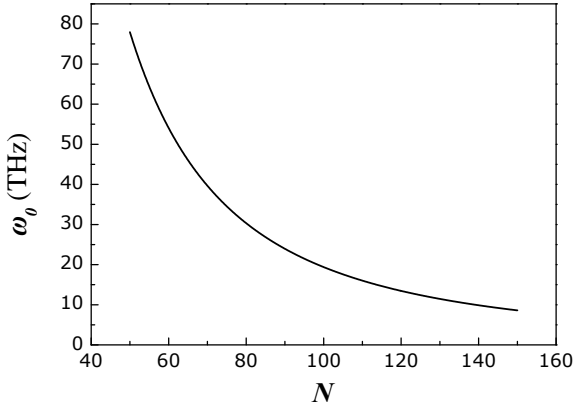}
  \caption{Frequency $\omega_0$ of the magnetic resonance of a 2D array of undoped CNS
 with 60\% surface coverage as a function of the number $N$ of turns of graphene sheet
 for a single CNS.}
\label{fig3}
\end{figure}

A lattice of CNS (miniaturized Swiss rolls) behaves as a uniaxial
magnetic metamaterial \cite{wiltshire_01} with an effective
magnetic permeability along the CNS axis provided by
\cite{pendry_99}

\begin{equation}
\mu_{eff} = 1 - \frac{F \omega^2} {\omega^2 - \omega_0^2 + i
\omega \gamma}, \label{eq:mu}
\end{equation}
where $F$ is the surface coverage of a single CNS within a single
unit cell. $\omega_0$ is the resonance frequency provided by
\begin{equation}
\omega_0 = c \sqrt{\frac{d}{2 \pi^2 r^{3} (N-1)}}
\label{eq:res_freq}
\end{equation}
with $c$ being the vacuum speed of light and $\gamma$ the loss
factor given by \cite{pendry_99}
\begin{equation}
\gamma = \frac{2 R_{s}} {r \mu_{0} (N-1)},\label{eq:gamma}
\end{equation}
where $\mu_0$ is magnetic permeability of vacuum and $R_{s}$ the
sheet resistance of a single layer of graphene. For undoped
graphene $R_{s}=200 \Omega$ \cite{cai}, whilst for doped graphene
can reach values as low as $R_{s}=11 \Omega$. \cite{de} Either for
doped or undoped graphene, these very low values of sheet
resistance ensure a very narrow magnetic resonance that becomes
narrower as the number of turns $N$ increases. Indeed, in
Fig.~\ref{fig2} we show the effective permeability $\mu_{eff}$ for
a 2D array of CNS with $F=0.6$. Each CNS contains $N=50$ turns of
graphene giving a magnetic resonance at $\omega_0=77.95$~THz with
FWHM of 3.2~GHz. The maximum value of the imaginary part of
$\mu_{eff}$ is above 12000 whereas the corresponding real part
varies from -6000 up to 6000. These extremely high absolute values
of $\mu_{eff}$ and its ultra-subwavelength nature ($\lambda / a =
212$) are responsible for the excellent imaging properties of the
CNS arrays shown below. The position of the magnetic resonance can
be tuned at a desired spectral window by varying the number of
turns $N$ of graphene layers, as demonstrated in Fig.~\ref{fig3}.

A note on the applicability of the formula (\ref{eq:mu}) is needed
for the case of the CNS arrays. Eq.~(\ref{eq:mu}) is proved for a
rolled-up meta-atom made of a perfectly conducting sheet. In order
to meet this requirement for a graphenic CNS, it should contain
enough turns $N$ of graphene so that the magnetic resonance
$\omega_0$ of Eq.~(\ref{eq:res_freq}) lies within the THz regime.
The dielectric function of graphene assumes very high absolute
values in this frequency range \cite{vakil,francescato} and can
therefore be safely considered as a perfect conductor. In the
optical regime, the above formula should be revised in order to
accommodate the inter- and intra-band transitions in graphene.
However, in this case, the magnetic activity of a CNS array would
deteriorate due to the high absorbance of graphene in the optical
regime induced by the above transitions. \cite{nikitin,simsek}

To draw a comparison with the state of the art in similar designs,
rolled-up metamaterials based on semiconductor microhelices and
operating at about 4~THz have been realized,
\cite{rottler,schwaiger} but the maximum reported value for the
real part of $\mu_{eff}$ is below 3, the corresponding FWHM about
1~THz, and the wavelength-to-structure ratio $\lambda / a \approx
7$. Clearly, the resonance features of $\mu_{eff}$ of the CNS
array (Fig.~\ref{fig2}) are superior to the above designs.

Next, we assess the imaging properties of a finite slab of the CNS
array studied in Fig.~\ref{fig2}. Due to the ultra-subwavelength
nature of the CNS array, the effective-medium approximation is
very accurate for our purpose, as has been demonstrated by direct
comparison of the effective-medium theory with exact numerical
calculations. \cite{demetriadou} In this context, a finite slab of
a 2D CNS array can be simulated as a homogeneous uniaxial medium
with $\mu_{xx}=\mu_{yy}=\mu_{\parallel}=1$ and $\mu_{zz}$ provided
by Eq.~(\ref{eq:mu}). In the calculations, we assume a slab of the
CNS array of Fig.~\ref{fig2} of 12.1~$\mu$m thickness. The
transmission coefficient for the magnetic field of a transverse
electric (TE) wave incident on a finite uniaxial magnetic slab of
thickness $d$ is given by \cite{wiltshire_07}
\begin{equation}
t = \Bigl[ \cos (k_{z} d) + \frac{1}{2}(\frac{\mu_{\parallel}
k_{\parallel}} {k_{z}} - \frac{k_{z}}{\mu_{\parallel}
k_{\parallel}}) \sin(k_{z}d) \Bigr]^{-1}, \label{eq:trans}
\end{equation}
where ${\bf k}_{\parallel}=(k_{x},k_{y})$ is the component of the
incident wavevector ${\bf k}_0$ which is parallel to the faces of
the slab and $k_{z}$ is the component normal to the slab, i.e.,
\begin{equation}
k_{z} = \sqrt{\mu_{\parallel} (k_0^2 - k_{\parallel}^{2} /
\mu_{z})}. \label{eq:k_z}
\end{equation}
For a wave decaying in free space, $k_{z}$ is an imaginary
quantity because $k_{0} < k_{\parallel}$. Nonetheless, within a
medium with $\mu_{\parallel} > 0$ and $\mu_{z} < 0$, $k_{z}$
becomes real corresponding to a propagating wave. This is the
typical operation of a hyperbolic electric or magnetic
metamaterial. When a free-space evanescent wave is incident on a
slab of a hyperbolic metamaterial, it is transported through the
slab without decay, or even with amplification if the imaginary
part of $k_{z}$ is negative (canalisation effect). The CNS array
studied here operates as a hyperbolic magnetic metamaterial in the
region of negative real part of $\mu_{eff}$ shown in
Fig.~\ref{fig2}. However, here we focus on a different regime,
namely around the frequency where $\mu_{z}$ becomes maximum,
approximating the theoretical limit of $\mu_{z}=\mu_{eff}
\rightarrow \infty$. In this case, we find from Eq.~(\ref{eq:k_z})
that $k_{z} \approx k_{0}$ (since $\mu_{\parallel}=1$) and all
incident waves (both far- and near-field) propagate through the
slab with the free-space wavenumber $k_{0}$.

\begin{figure}[h]
\centering
  \includegraphics[width=8cm]{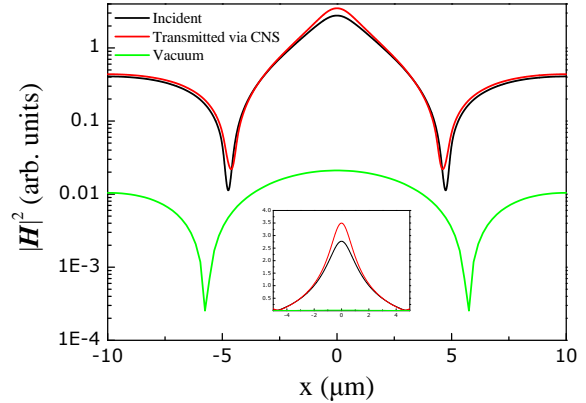}
  \caption{Imaging via a slab of CNS. Incident (input) and transmitted (output) magnetic field via a
  finite slab of CNS, i.e., those considered in Figs.~\ref{fig2} and \ref{fig3} with hexagonal cross
  section (of 10$\mu$m side) and 12.1$\mu$m thickness. The source of the magnetic field is a current loop (magnetic dipole) of
  0.3$\mu$m radius and is placed at a distance of 0.5$\mu$m from the left face of the slab. We also show
  the magnetic field in the absence of CNS slab, i.e., at a vacuum distance of 12.9$\mu$m from the source loop.
  The magnetic field is calculated along a line joining the middle points of two parallel sides of the hexagonal cross section.
  The incident magnetic field is calculated at the left face of the CNS slab whilst the transmitted field
  is calculated at a plane parallel to the slab at 0.3$\mu$m from its right face.}
\label{fig4}
\end{figure}

Figure~\ref{fig4} depicts the image of the magnetic field of a
magnetic dipole (source), specifically, a current loop of
0.3~$\mu$m radius. In the same figure, we show for comparison the
magnetic field at the same distance from the source loop, in the
absence of the CNS slab. Evidently, the slab of the CNS array
transfers perfectly (with some amplification as well) the image
from the magnetic dipole. Without the CNS slab, the image decays
by more than two orders of magnitude. We note here that the value
of 12.1~$\mu$m for the thickness of the slab has been chosen so as
to eliminate the reflection from the surface of the slab. Namely,
for this thickness, $k_{z} d = n \pi$ ($n$ is an integer) the
transmission factor $t$ in Eq.~(\ref{eq:trans}) becomes unity. Of
course, this is possible because $k_{z}$ inside the CNS slab
becomes real (whereas in free space it is purely imaginary, as
discussed above)

\begin{figure}[h]
\centering
  \includegraphics[width=8cm]{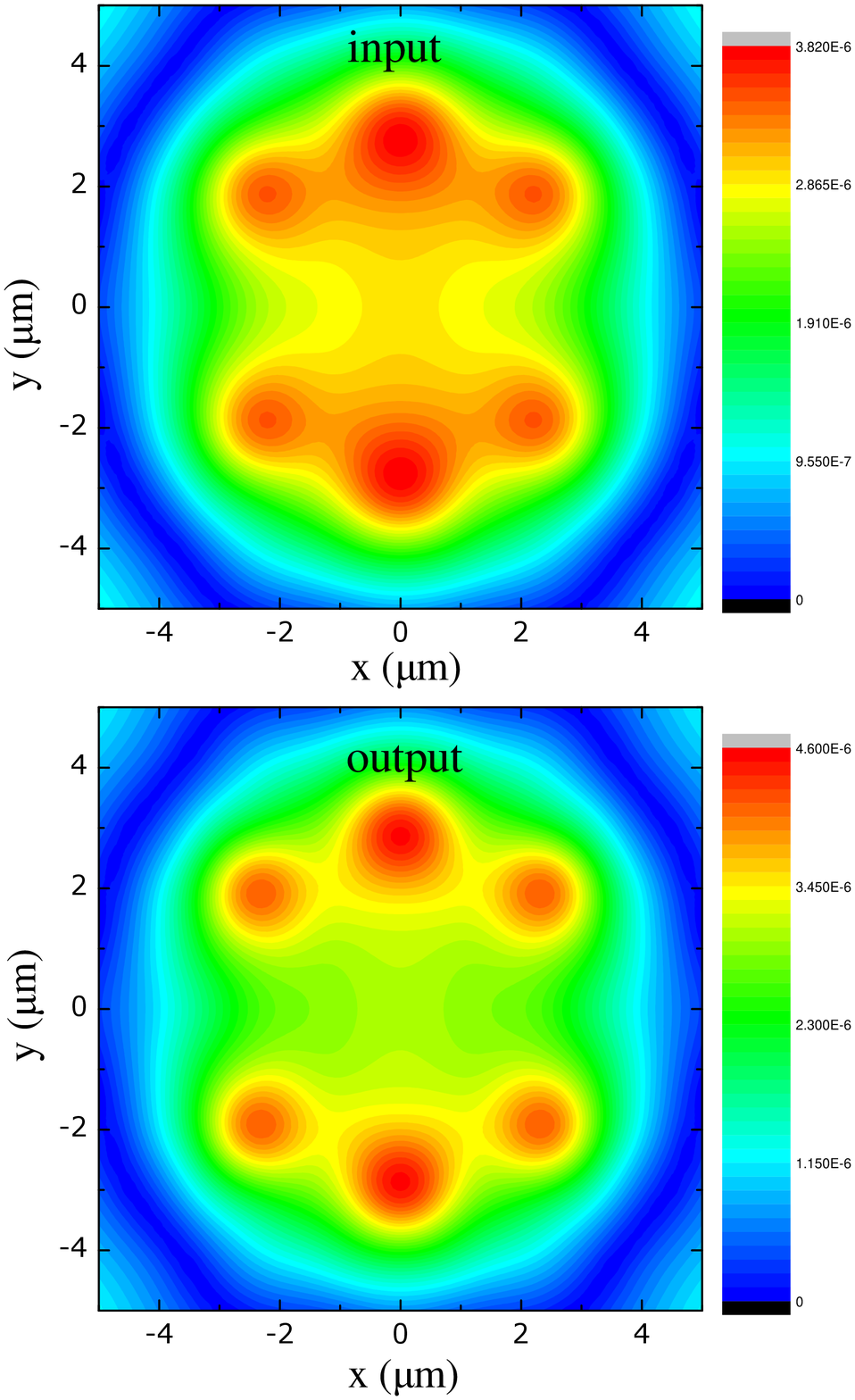}
  \caption{Imaging via a slab of CNS of multiple sources. Incident (input) and transmitted (output) magnetic field via
  the slab of CNS considered in Fig.~\ref{fig4}. The input
  magnetic field is produced by six current loops (same as that of
  Fig.~\ref{fig4}) placed within a plane parallel to the faces of
  the slab at positions (x,y)=(0,3$\mu$m),(2.5$\mu$m, 2 $\mu$m), (2.5$\mu$m, -2 $\mu$m), (0,-3$\mu$m), (-2.5$\mu$m, -2 $\mu$m), (-2.5$\mu$m, 2 $\mu$m), at a distance of 0.5$\mu$m from the left face of the slab.
  The incident magnetic field is calculated within the left face of the CNS slab whilst the transmitted field
  is calculated within a plane parallel to the slab at 0.3$\mu$m from its right face.} \label{fig5}
\end{figure}

In Fig.~\ref{fig5} we demonstrate the imaging functionality of the
CNS slab when multiple sources constitute a probe image incident
on the same slab as in Fig.~\ref{fig4}. Obviously, the image of
the multiple sources is very efficiently transferred through the
CNS slab, resulting in a clearer image of the sources due to the
slight amplification of the magnetic field. In other words, the
CNS slab not only operates as a THz endoscope of the magnetic
field, but also offers a better resolution of the reconstructed
image due to the slight amplification of the incident near field.

\begin{figure}[h]
\centering
  \includegraphics[width=8cm]{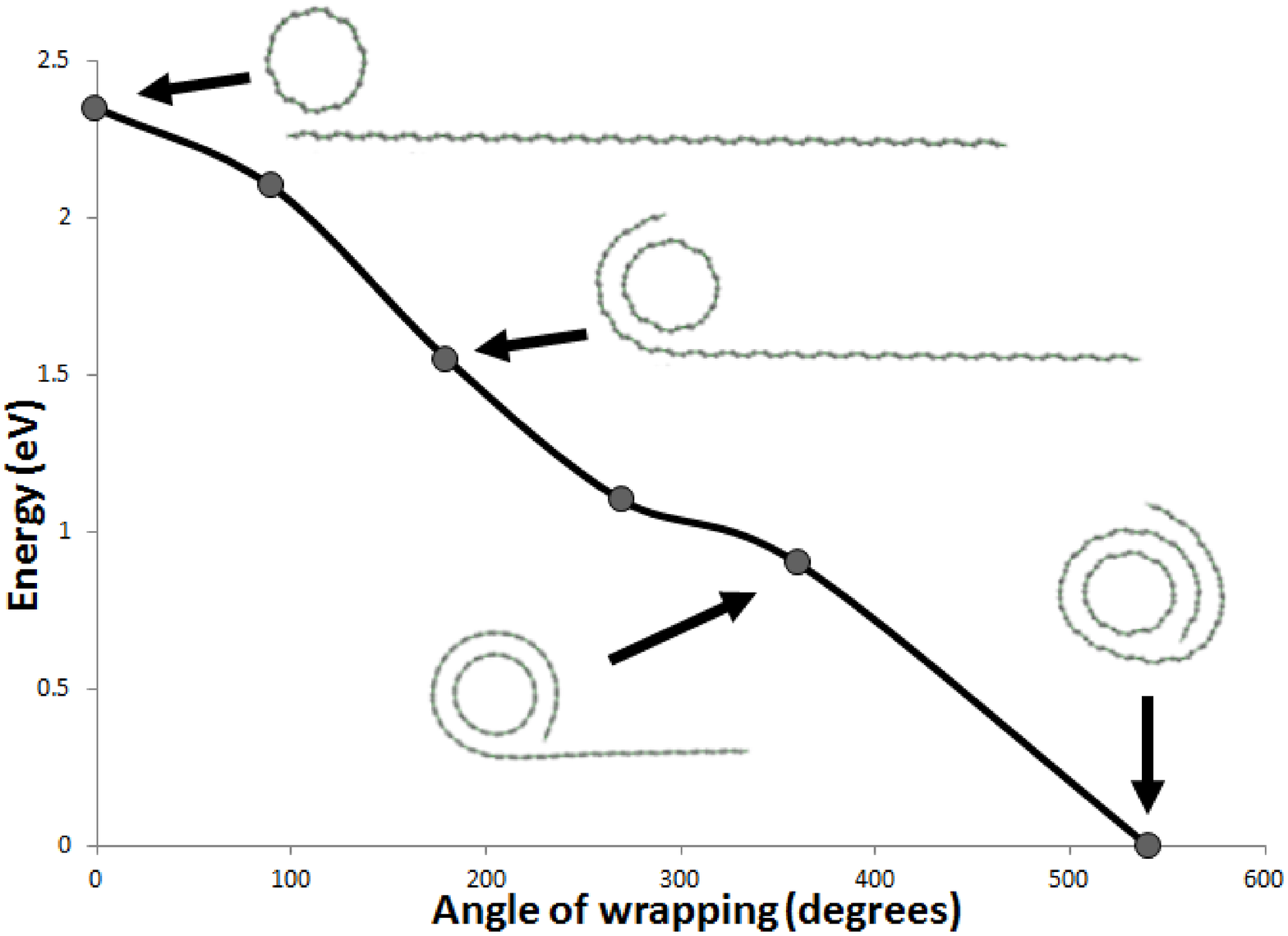}
  \caption{Energy drop during wrapping of a graphene nano-ribbon around
  a (15,15) single-wall carbon nanotube. Insets shows the partially formed
  carbon nano-scroll for different levels of wrapping.} \label{fig6}
\end{figure}

We note that we have considered an ordered array of CNS operating
as uniaxial homogeneous medium. An isotropic homogeneous medium
can be obtained if the CNS are not periodically positioned in
space, for example, in random-packing fashion with arbitrary
orientations of the CNS axis. In this case, the effective
permeability will be isotropic (scalar quantity) and it will be
again provided by Eq.~(\ref{eq:mu}). Moreover, when the CNS
magnetic metamaterial is blended with randomly positioned and
oriented long metallic nanorods which also exhibit electric
response in the THz regime, a negative refractive index can be
achieved.

We now address the issue of how an array of carbon nanoscrolls may
form. There are two key facts that suggest the feasibility of this
formation. First, vertical arrays of carbon nanotubes can be grown
on various substrates with good control over the positions and
inter-tube distance.\cite{heer,bennett_adm06} In turn, these
arrays can serve as seeds for the formation of a lattice of CNS
with graphene sheets wrapped around the nanotubes. The wrapping
process has been described in previous computational
studies~\cite{zhang_apl10,patra_acsn11,cheng_phe12} based on
calculations with empirical force-field potentials. In the
following we present the details of graphene wrapping around a
single-wall carbon nanotube (SWCNT) using first-principles
calculations.

The results were obtained with the density-functional theory code
VASP,\cite{VASP} using projector-augmented waves,\cite{PAW} an
energy cutoff of 350 eV for the plane-wave basis, and a
generalized-gradient exchange-correlation functional.~\cite{PBE}
We examined the wrapping of zig-zag graphene nanoribbons (GNR)
with H-passivated ends around (6,6), (9,9) and (15,15) arm-chair
SWCNTs, as well as the rolling of arm-chair GNRs around (9,0),
(13,0) and (16,0) zig-zag SWCNTs. In the supercells used, one
dimension is small and corresponds to the size of the unit cell of
the carbon nanotube. Sampling of reciprocal space employed an
8$\times$1$\times$1 $k$-grid with the first component related to
the small dimension of real space.

Figure~\ref{fig6} shows the decrease of energy when a a zig-zag
GNR wraps around a (15,15) SWCNT. The monotonic drop of energy as
the wrapping angle increases indicates that the CNS is formed
readily for this particular nanotube. Similar results are obtained
for rolling around the large diameter (16,0) SWCNT. In contrast,
wrapping around the (6,6) and (9,0) nanotubes is not an exothermic
process as the strain energy component becomes prohibitively large
when the SWCNT diameter is small. The cases of (9,9) and (13,0)
SWCNTs seem to be close to the threshold diameter that favors CNS
formation. Notwithstanding the practical challenges, the
computational results demonstrate the feasibility of growing
arrays of carbon nano-scrolls around ordered seeds of carbon
nanotubes that are not too small. Finally, we note that the
magnetic response of a single SWCNT is irrelevant in our case
since the resonance in $\mu_{eff}$ is very weak than that of a
single CNS and lies at much higher frequencies. \cite{pendry_99}

In conclusion, we have shown that one of the many conformations of
single-layer graphene, namely, carbon nanoscrolls act as a
magnetic metamaterial in the THz where naturally occurring
magnetic materials are particularly scarce. Thanks to the small
size of the carbon nanoscrolls compared to the operating THz
wavelength (two orders of magnitude smaller), their spiral shape
and the low sheet resistance of graphene, carbon nanoscrolls
exhibit unrivaled imaging properties for magnetic sources in the
THz regime. One of the possible routes to realize an array of
carbon nanoscrolls is by self-folding of a single layer of
graphene around a carbon nanotube, ordered arrays of which have
already been realized in the laboratory. \cite{heer,bennett_adm06}

\bibliographystyle{rsc}


\end{document}